\documentclass[10pt,letterpaper,twocolumn]{article}

\usepackage[letterpaper,margin=1in]{geometry}
\usepackage{times}
\usepackage{cite}
\usepackage{url}
\usepackage{graphicx}
\usepackage{xcolor}
\usepackage{xspace}
\usepackage{cite, url}
\usepackage{hyperref}
\usepackage{balance}
\hypersetup{
  colorlinks=true,      
  linkcolor=blue,       
  citecolor=blue,       
  filecolor=cyan,       
  urlcolor=blue         
}

\usepackage[bf,small]{caption}

\usepackage{titlesec}
\titlespacing*{\section}{0pt}{*1}{*1}
\titlespacing*{\subsection}{0pt}{*1}{*1}
\titlespacing*{\subsubsection}{0pt}{*1}{*1}
\titlespacing*{\paragraph}{0pt}{*1}{*1}
\def\eg{e.g.,\xspace}
\def\ie{i.e.,\xspace}

\title{\bf Treehouse: A Case For Carbon-Aware Datacenter Software}
\author{
Thomas Anderson$^{1}$, Adam Belay$^{2}$, Mosharaf Chowdhury$^{3}$, Asaf Cidon$^{4}$, and Irene Zhang$^{1,5}$\\
{\normalsize \vspace{6mm}$^{1}$University of Washington, $^{2}$MIT, $^{3}$University of Michigan, $^{4}$Columbia University, $^{5}$Microsoft Research}
}
\date{}

\newcommand{\tom}[1]{\textcolor{red}{(#1)}}

\newcommand{\function}{$\mu$function\xspace}
\newcommand{\functions}{$\mu$functions\xspace}

\begin{document}
\maketitle
\begin{abstract}
The end of Dennard scaling and the slowing of Moore's Law has put the energy use
of datacenters on an unsustainable path.  Datacenters are already a significant
fraction of worldwide electricity use, with application demand scaling at 
a rapid rate. We argue that substantial reductions in the carbon intensity 
of datacenter computing are possible with a software-centric approach:
by making energy and carbon visible to application developers on a fine-grained basis,
by modifying system APIs to make it possible to make informed trade offs
between performance and carbon emissions, and by raising the level of 
application programming to allow for flexible use of more energy efficient
means of compute and storage. We also lay out a research agenda for
systems software to reduce the carbon footprint of datacenter computing.
\end{abstract}


\section{Introduction}
The pressing need for society to address global climate change has
caused many large organizations to begin to track and report their
aggregate greenhouse gas emissions, both directly caused by their
operations and indirectly caused through energy use and by
supply chains~\cite{ghg}.  However, there are no standard
\emph{software} mechanisms in place to track and and control emissions
from information technology (IT).  This lack of visibility is
especially acute where multiple applications share the same physical
hardware, such as datacenters, since carbon emissions today can only
be accounted for at the server or processor chip level, 
not at the software and application level.

\begin{figure}[t]
  \centering
  \includegraphics[width=\columnwidth]{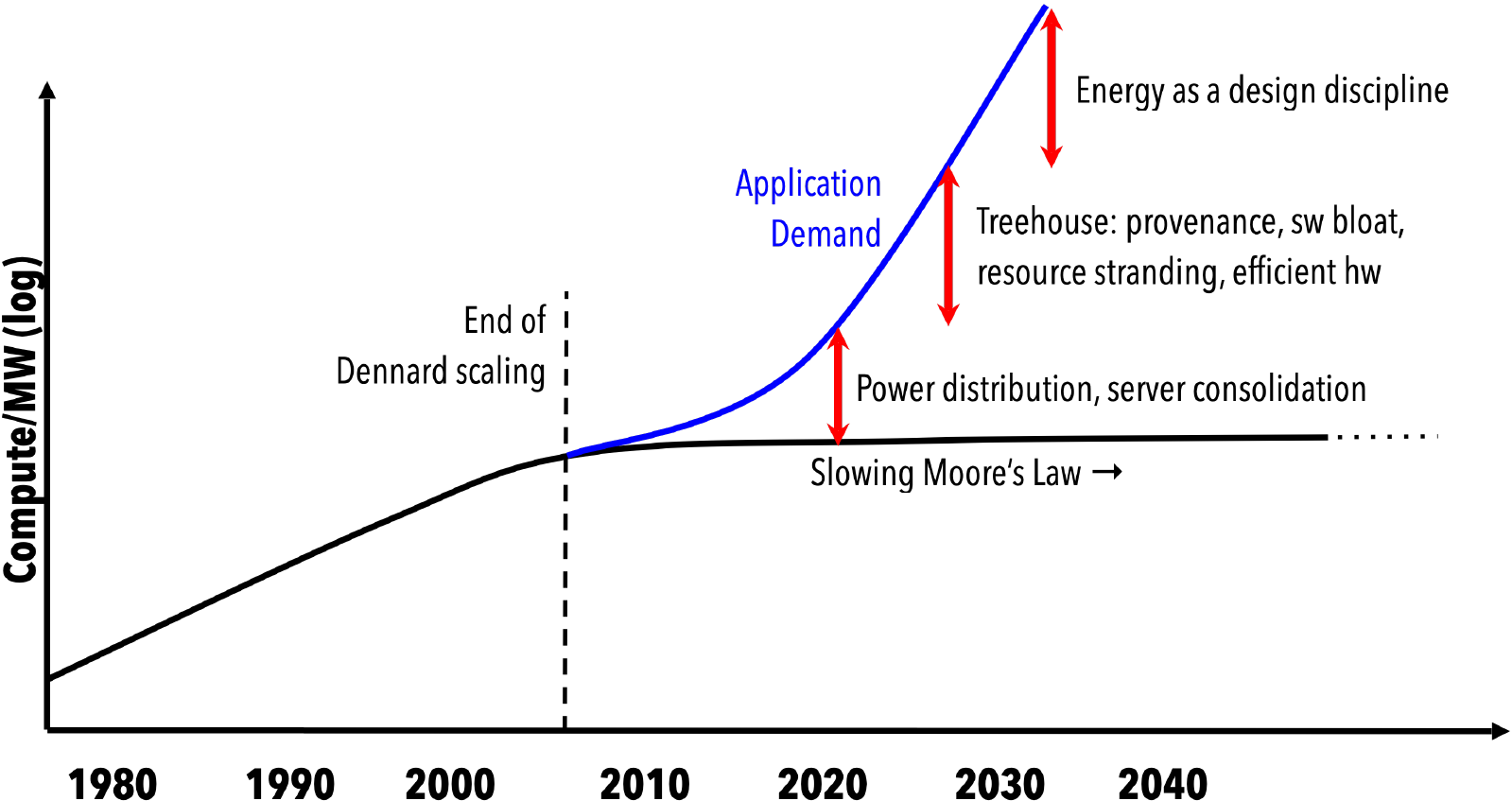}
  \caption{Application demands for energy is growing faster than energy efficiency improvements we can achieve.
    Treehouse takes a software-centric approach to reduce this gap.}
  \label{fig:trends}
\end{figure}

In aggregate, datacenters represent a large and growing source of
carbon emissions; estimates place datacenters as responsible for 1-2\%
of aggregate worldwide electricity
consumption~\cite{jones2018stop,pearce2018energy}.  Given
rapidly-increasing demand for computing and data
analysis~\cite{recalibrate-dc-energy,vahdat-sigcomm}, continual
improvements are needed in the carbon efficiency of computing to keep
the climate impact of computing from
skyrocketing~\cite{jones2018stop,pearce2018energy,pesce}.  The end of
Dennard scaling means that exponential improvements in energy
efficiency are no longer an automatic consequence of Moore's Law.
Over the past few years, various technologies have been introduced to
improve matters---for example, server consolidation and improvements
in power distribution.  However, these steps will not be enough going
forward (Figure~\ref{fig:trends}).

For cloud datacenter operators, a popular option is to construct
datacenters in locations with inexpensive, renewable power generation.
Although a step forward, this is unlikely to be a complete solution
for several reasons. First, hardware manufacturing, assembly and
transportation, as well as the construction and maintenance of the
datacenter itself, are all energy and greenhouse gas intensive. In
fact, chip manufacturing alone is responsible for about a third of the
lifecycle greenhouse gas emissions of a modern
datacenter~\cite{chasingcarbs}.  Second, edge computing---placing
computing near customers---is increasingly popular as a way to improve
application responsiveness; these smaller scale datacenters are often
located in or near cities without access to dedicated green power
sources.\footnote{For example, about a half acre of solar panels, plus
  batteries, are needed to fully power a single 24x7 server
  rack~\cite{landuse}.}  Power is often much slower to provision than
other parts of IT; for example, provisioning interstate power lines to access remote
green energy often requires many years of advance planning. 
Finally, many companies continue to operate
their own on-premise datacenters; any solution must work for those
deployments as well.  

We propose Treehouse, a project whose goal is to build the foundations for a new
software infrastructure that treats energy and carbon as a first-class resource, alongside traditional
computing resources like compute, memory, and storage. 
Today, developers have almost no way to know how their engineering decisions affect the climate.
The goal of Treehouse is to enable developers and operators to understand and reduce greenhouse
gases from datacenter sources. 
We target all datacenter environments, including cloud, edge computing, and on-premise environments.

We identify three new foundational abstractions necessary to enable developers to optimize their
carbon footprint: (1) energy provenance, a mechanism to track energy usage, (2) an interface for expressing applications'
service-level agreements (SLAs) to allow operators to trade off performance and carbon consumption, and
(3) \functions, a fungible fine-grained unit of execution that enables more efficient hardware utilization.
We also lay out a research agenda to develop mechanisms for reducing carbon footprint by: (1) reducing
software bloat, (2) interchanging computational resources, (3) interchanging memory resources, 
and (4) energy-aware scheduling policies.

Beyond our direct research agenda, we hope our efforts can inspire the broader
software systems community to focus much more on datacenter carbon reduction.

\section{Foundations for Energy-Aware Datacenter Software}

Application developers today have few tools at their disposal to write energy and carbon-efficient
applications. First, they have no good way to account for the amount of carbon their applications are emitting.
In addition, it is not clear what the carbon implications would be of particular design choices
(\eg shifting their application from a dedicated server to a shared server, or moving their storage from disk to flash).
While many cloud users do optimize for lower cloud costs, cost does not equate to energy usage.
For example, while an HDD is much cheaper to run than an SSD, it is far more energy intensive.
Similarly, for computationally intensive applications, FPGAs can often provide only a small 
integer factor speedup relative to CPUs, but a factor of 10-70 improvement in energy efficiency~\cite{avocados}.

Second, from the standpoint of the operator (\eg the cloud provider or a devops engineer in an on-premises data center),
even if they have some understanding of the energy consumption of particular
hardware resources (\eg servers), reducing the carbon footprint of a workload will often reduce performance.  The datacenter operator does not typically know
when it would be appropriate to make that tradeoff, and so the common practice is to optimize 
infrastructure energy use only when it would have negligible impact on performance, 
regardless of whether the performance matters to a particular application.

Third, software applications today are typically provisioned in a static set of bundled resources, which make it difficult
to optimize for lower energy usage. For example, virtual machines or containers typically come pre-allocated with a set of CPU cores, memory capacity, and network and disk bandwidth. As modern datacenter applications typically
exhibit bursty and unpredictable patterns at the microsecond-scale, this bundling of resources causes applications to be
inefficient and energy-wasteful.

In this section, we introduce a set of abstractions that we believe will lay the 
foundations for solving these problems,
to allow developers to track and optimize the energy and carbon footprints of their applications.


\subsection{Energy Provenance}

In order to track and account for carbon emissions at the software
level, we need the ability to measure the \emph{energy provenance}
of each application. We use the term provenance to denote both
the direct and indirect energy usage of a particular application.
For example, an application not only directly consumes energy when it is running
user-level code, but it also consumes energy in the operating system, in storage devices,
and in the network interface and switches along its path when it is communicating 
with a remote server, as well as the energy used on its behalf at the remote server.

Since it is difficult to directly measure the lifecycle energy provenance of 
individual applications through hardware mechanisms alone, we believe it will be
necessary to construct
a supervised machine learning model
to estimate the energy provenance of the application, given its resource usage.
The input (or features) of the model
will be metrics that are easily measured in software, including the network bandwidth (for switches and network interface cards),
bytes of storage and storage bandwidth (for memory and persistent storage)
and accelerator cycles, as well as the type and topology of hardware the application runs on.
The model could be trained and validated by carefully measuring in a lab environment how these performance
metrics affect system-level energy usage.
Armed with accurate single-node energy provenance estimates, we plan to annotate data center
communication, such as remote procedure calls (RPCs), much as cloud providers
annotate RPCs with debugging information today~\cite{xtrace}. These lifecycle per-application
energy estimates, combined with estimates of the carbon intensity of power generation in
each location, would give developers the needed visibility into the impact of their design 
decisions. This is a necessary first step to enlisting the
developer community in achieving computational energy and carbon efficiency.


\subsection{Exposing Application-Level SLAs}

Another barrier to energy-efficient computing is that optimizations that improve
energy efficiency often hurt performance. Disabling processor boost mode; moving
less frequently used data from high power memory to much lower power
non-volatile memory or solid-state storage; turning off underutilized memory
chips; moving computation from power-hungry general purpose processors to more
efficient dedicated hardware accelerators; powering down a fraction of the
network core when it is not needed---these steps save energy but very often come
at the cost of worse system and application performance.

For application code, provided we address energy provenance, the application
developer can decide on the right tradeoff that meets user performance
expectations in the most energy-efficient manner possible.  These types of
optimizations are harder for systems code because it currently lacks any direct
knowledge of application intent.  Traditionally, system designs have been
evaluated in terms of response times and throughput, but to achieve this
designers have been willing to use all available resources regardless of the
energy cost. Thus, while these designs are often optimal for performance, they
sacrifice carbon efficiency.

To address this challenge, we aim to provide a way for application developers to
convey to systems code their tolerance (and/or desire) for energy-saving
optimizations.  This is the equivalent of eco-mode when driving a car.  Together
with provenance data to track the energy impact of using different resources,
the system designer and operator can make informed choices as to how to schedule
and place workloads. Once system code can optimize its behavior along the
energy-performance Pareto curve, application developers can make informed
choices to meet their users' carbon reduction goals.

To this end, we believe we need to develop a new interface to expose application-level 
performance constraints (Service Level Agreements, or SLAs) to systems software. This will enable a new class of 
energy-aware systems-level optimizations.
For highly latency-sensitive operations, it may still make sense to use the highest-performance
solutions, even at high energy cost.  But where there is available slack in user expectations,
we can use that flexibility to choose the most energy-efficient solution consistent with
meeting user needs.

\begin{figure}
  \centering
  \includegraphics[width=\columnwidth]{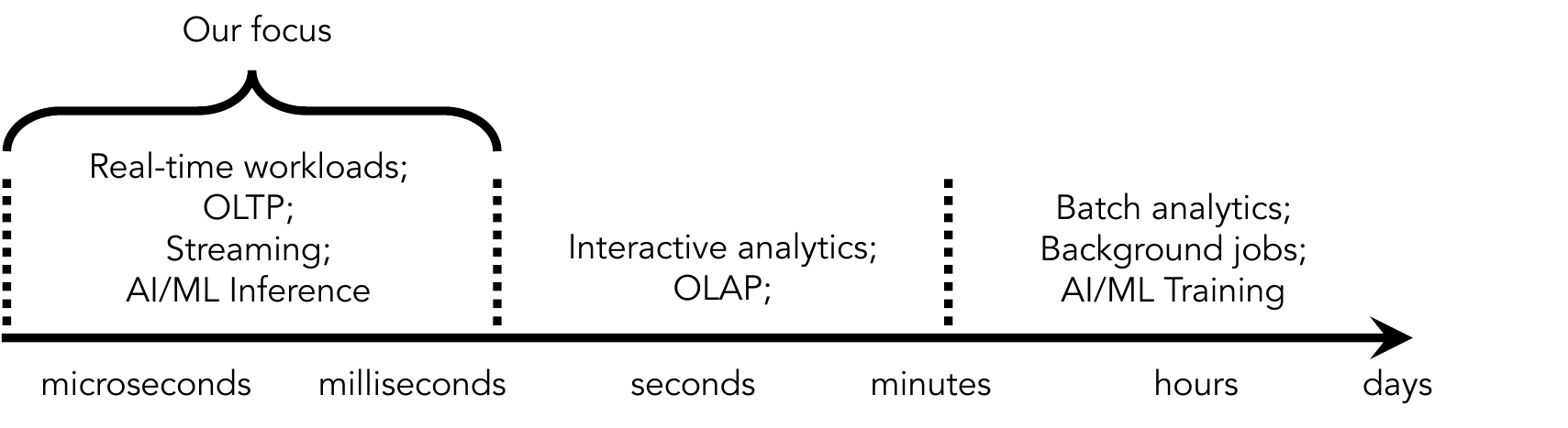}
  \caption{Treehouse focuses on reducing the carbon footprint for tasks with sub-second SLAs. 
Due to these latency constraints, the energy optimizations available are those within the 
same datacenter. Prior work has considered relocating batch jobs (\eg analytics) to 
datacenters with greener sources of power, or to greener periods of the day.}
  \label{fig:SLA}
\end{figure}

There is a large body of work on shifting long-running
batch jobs (\eg MapReduce-style analytics) to a cleaner sources of energy~\cite{liu2011greening,chen2012energy,mashayekhy2014energy,cheng2015towards,chen2010compress,krioukov2011integrating,kim2016data}. 
These type of tasks are the extreme end of the SLA spectrum (depicted in Figure~\ref{fig:SLA}), and typically operate at time scales of hours or even days. This provides
enough slack to shift them to geographically-remote datacenters or to different times of the day, 
to take advantage of spatial or temporal availability of clean energy (\eg wind and solar).
Our focus is on energy optimizations that can also apply to applications
with much tighter SLAs, at the millisecond and even single-digit microsecond scale.
For these applications, it isn't feasible to move the work to remote datacenters or to periods of 
off-peak electricity generation. 


\subsection{\functions}

Despite the fact that many applications have highly dynamic resource
usage, cloud applications today are often provisioned for peak resource usage in
coarse-grained and static ways.  For example, a virtual machine,
container, or even serverless compute engine will be provisioned
statically, with, say, 4 cores, 32~GB of memory, etc., for seconds,
minutes and hours at a time, while application demand varies at much
finer time-scales.

This leads to a high degree of resource stranding---compute, memory,
and storage that is only lightly utilized, but which cannot be used
for other applications.  Although many hardware devices have low power
modes, these are only of partial benefit.  Even at low load, power
consumption is often half of the high load case~\cite{energyprop}, in
addition to the environmental impact of fabricating devices that
on average sit idle.  Power efficiency per unit of
application work is maximized when system software keeps resources
fully utilized.

Further, the most energy efficient option is to avoid doing work that
wasn't needed in the first place.  Existing datacenter software stacks
are bloated, with layers of functionality added over time and kept for
programmer speed and convenience rather than refactored down to their
essential purpose.  In the old era of Dennard scaling, inefficient
layering could be addressed with time---every year, faster and more
energy-efficient computers would become available to hide the impact
of software bloat.  With the end of Dennard scaling, however, keeping
old, inefficient software layers adds up.

We believe we need a new abstraction to address both software bloat
and resource stranding.  First, we need a lightweight way to provision
resources at much finer time scales, choosing the most energy
efficient option that meets each application's SLA. Second, to achieve
high utilization, we need to aggregate application resource demands
more effectively.

\paragraph{A New Abstraction for Fungible Compute}
Modern datacenter applications are distributed at extremely fine
granularities. For example, each user-facing HTTP request received by
Facebook or Twitter spawn requests to dozens of microservices that
lead to thousands of individual RPCs to servers.  As
datacenter networks get faster and in-memory microservices become more
efficient (e.g., by using kernel-bypass), datacenter servers can
increasingly process and respond to requests in
microseconds~\cite{killermicroseconds}.

To accommodate microsecond-scale datacenter applications, we need a
new programming model with fine-grained resource allocation and low
provisioning overheads.  It must be efficient enough to make
adjustments at the microsecond-scale, so it can respond to sudden
workload changes~\cite{snap,perfiso}.  Reflecting its
microsecond scale, we call this abstraction for general-purpose
fine-grained application provisioning {\em microfunctions}.
Microfunctions represent a large enough time scale to do useful work
(i.e., a few thousand cycles), while being fine-grained enough to
balance resource usage quickly as shifts in load occur.  

We plan to use an RPC-based API for \functions, including an interface for the
user to define SLAs, as well as an energy or carbon budget. We also see foresee
opportunities to further increase efficiency through computation shipping
between \functions, allowing us to improve locality and reduce data
movement~\cite{kayak, splinter}. Dynamically deciding when to move data or computation
will also enable new efficiency vs. performance tradeoffs. Building upon the
recent trend toward microservices, we envision that full applications can be
constructed by partitioning their components into fine-grained units and
running them as independent \functions.

FaaS (Function-as-a-Service) or serverless frameworks, such as AWS
Lambda~\cite{aws-lambda} share a similar notion by allowing developers
to express their jobs and get billed at the granularity of individual
function invocations. However, FaaS still operates on top of
statically allocated resource containers, making it difficult to bin
pack the right combination of functions---in the face of variable
resource usage---to achieve high utilization.  Some cloud providers
compensate by overcommitting functions to containers, but this leads
to inconsistent per-function performance~\cite{behindserverless}. In
addition, FaaS suffers from software bloat and high function startup
times. The ``cold start'' problem, in particular, can cause FaaS to
take hundreds of milliseconds or more to invoke a
function~\cite{shahrad2020serverless}. This timescale is many orders
of magnitude too coarse to achieve balance during fine-grained shifts
in resource demand, while incurring far higher energy overhead than is
necessary.
Finally, FaaS is only designed to operate on a specific type of
compute and memory (namely, CPU and DRAM), and cannot take advantage
of more energy efficient options such as accelerators (\eg GPUs,
FPGAs, NICs) and heterogeneous forms of memory (\eg persistent
memory).

Our goal for \functions is to provide a \emph{lightweight} function
abstraction, which is \emph{decoupled from any static grouping of
  resources}, such as a container or a VM. Instead, we aim to make
\functions completely fungible, consuming resources on-demand as they
are needed, with the ability to run on heterogeneous computing
resources.

\paragraph{Micro-Second Scale Performance}

In order to exploit fine-grained variations in resource usage and concurrency, we plan to
support microsecond-scale invocations of \functions, an improvement of several
orders of magnitude over existing serverless systems. We must tackle two
research challenges to spawn \functions this quickly.

First, the cold start problem must be addressed to speed up invocations on
machines that have not recently executed a particular function. One barrier is the
high initialization cost of existing isolation mechanisms. For example, even
after sophisticated optimizations, Amazon's Firecracker still requires at least 125 milliseconds to
start executing a function environment~\cite{firecracker}. 

Second, we must ensure that \function invocations themselves can start
extremely quickly. A major barrier to fast function
invocation in existing FaaS systems is that they rely on inefficient RPC protocols
built on top of HTTP.  
In addition, existing FaaS systems require a complex tier of dedicated load
balancing servers~\cite{firecracker}, which leads to significant delays. 

\paragraph{Resource Disaggregation}

Resource disaggregation poses a solution to the fixed bundling
of resources (in servers, virtual machines or containers).  While microsecond resource
allocation helps to minimize the resources stranded by overprovisioning, it does not
solve the problem of bin packing application resource allocations
onto servers, leaving some resources still stranded.  Disaggregating resources
reduces resource stranding at the cost of added latency. For applications whose SLAs
are designed to tolerate slightly longer latencies, disaggregation enables 
the system to allocate exactly the amount of compute, memory and storage
each application requires at the moment, from a shared pool. This allows idle resources
to be powered off to save energy without compromising application-level SLAs.

There has been some progress on disaggregating resources, particularly on disaggregating storage~\cite{ebs,s3,alibaba-disagg}.
However, some resources, such as memory and CPU, are still primarily consumed locally on monolithic servers.
While there is a large body of research on trying to disaggregate these resources~\cite{remote-regions,infiniswap,AIFM,app-performance-disagg-dc,farmemory-throughput,LegoOS,googledisagg}, significant challenges remain for real-world adoption, including: 
security~\cite{rdma-security}, isolation~\cite{justitia:nsdi22}, synchronization~\cite{ma2020asymnvm} and fault tolerance~\cite{hydra:arxiv19}.
These challenges are exacerbated especially in low-latency (\ie microsecond-scale) settings that
are our focus. 

\begin{figure}
  \centering
  \includegraphics[width=\columnwidth]{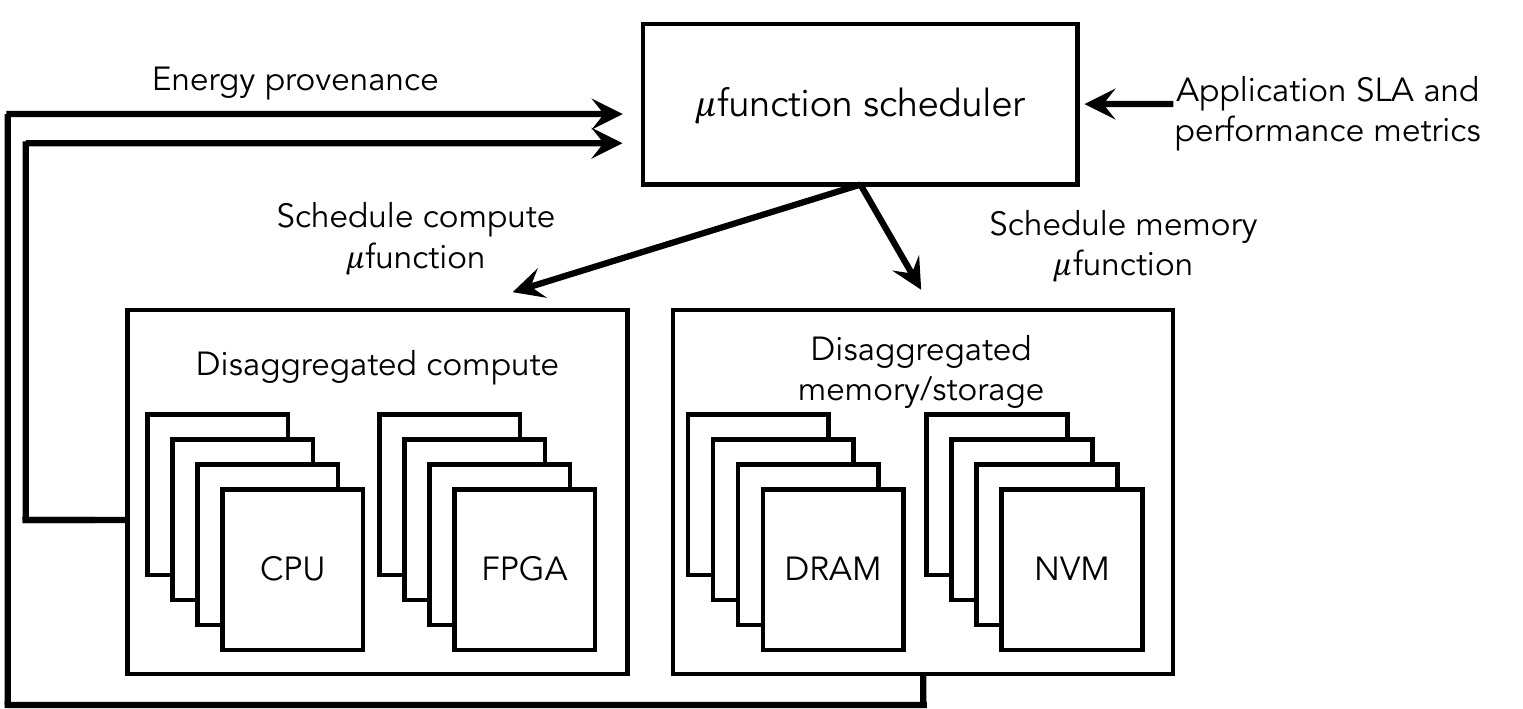}
  \caption{Depiction of the Treehouse scheduler. The scheduler takes as its input the energy provenance of each function, the state of the different hardware resources as well as the function's SLA. It then schedules the functions in the most energy-efficient way, while still meeting their SLAs, across the different clusters of disaggregated resources.}
  \label{fig:treehouse-scheduler}
\end{figure}

\paragraph{Design Questions}

A key design question is whether to build \functions on top of Linux, and whether
\functions need to be able to support POSIX. While running \functions on top of Linux may
make it easier for existing applications to transition to \functions, it comes at a high cost.
In particular, Linux adds significant overhead to I/O operations, and it is not the natural interface
for writing a distributed application across disaggregated resources.
We plan to pursue in parallel both research directions:
(a) incrementally adapt Linux to be more lightweight,
as well as (b) pursue a clean-slate non-POSIX OS design.
We describe these efforts in Section~\ref{sec:bloat}.

\subsection{Summary}

To conclude, our three foundational abstractions would allow developers to define \functions that can
operate on fungible resources at microsecond time-scales. Developers would define
SLAs for these \functions, allowing the cloud operator to navigate the energy-performance Pareto curve.
Finally, the energy provenance of these \functions would be tracked and accounted for at all times.

This process is depicted in Figure~\ref{fig:treehouse-scheduler}, where the Treehouse scheduler (described in \S\ref{sec:scheduler})
collects as input the energy provenance and the SLA of the \functions, and schedules them on the
resource at a time that would still meet their SLA while minimizing overall energy usage.

\section{Research Agenda}

We now describe a specific agenda that builds upon the Treehouse foundational abstractions to reduce
datacenter energy consumption, by allowing software systems to make energy-aware decisions.


\subsection{Minimizing Software Bloat}
\label{sec:bloat}

Inefficient software layers can be found in system-level building blocks shared across 
applications, including data movement, data (un)marshalling, memory allocation, and
remote procedure call handling.  In a cluster-wide profiling study at Google, it was found that
these common building blocks consume about 30\% of all cycles; the Linux kernel,
including thread scheduling and network packet
processing, consumes an additional 20\% of all cycles~\cite{profile_warehouse}.
In other words, shared software infrastructure is significant enough to account
for almost half of all CPU cycles available in a typical datacenter. 

We propose two steps to address software bloat. 
The first step is to continue optimizing
the many layers of the IT software stack that we have inherited.  Many
of these layers were designed for systems where I/O took milliseconds
to complete. We need a fundamental redesign of the
software stack for fast I/O (networking and storage) devices.

One direction is to use Linux as a control plane for backward compatibility,
but allow applications efficient direct access to I/O~\cite{arrakis,IX}.
Widely-used bypass technologies include RDMA and DPDK~\cite{dpdk} for network bypass 
as well as Optane and SPDK~\cite{spdk} for storage bypass. Although more work is needed
to understand how best to integrate these technologies with the kernel, studies have shown that
operating system overheads can be slashed while still providing traditional kernel functions
such as centralized scheduling, file system semantics, and performance 
isolation~\cite{strata,shenango,zhang2019i}. A complementary approach is to move
user-defined functions written in a type-safe language into the Linux kernel, 
to allow customization closer to the hardware~\cite{bento,bpf-storage,enberg2019partition,ghigoff2021bmc}.



A longer-term solution is to offload parts of 
the data path to more powerful and lower-energy I/O hardware.  For
example, both Amazon and Microsoft Azure offload to hardware the packet re-writing
needed for cloud virtualization~\cite{flexnic,azure-smartnics}.
This minimizes the energy cost of the added abstraction.  
We need to extend this approach to other layers of the systems stack
to truly reduce the software energy drain from management systems.
For example, we are designing an open-source, reconfigurable hardware 
networking stack to reduce energy use of frequently used operating system 
and runtime functions.

Ultimately, we believe we will need a new energy-optimized operating system kernel and runtime system
for datacenters architected to take advantage of energy-efficient hardware acceleration.  This
may be either as a clean-slate design or by incrementally replacing parts of the Linux 
kernel~\cite{secure}.  By raising the level of abstraction from POSIX to \functions, we make it easier
to support these more radical designs.

\subsection{Interchangeable Compute}
Datacenter applications are often designed to take advantage of a specific type of compute engine.
Traditional applications typically assume they are running on CPUs, while many machine learning 
applications rely on accelerators like GPUs, TPUs, and FPGAs, with new options emerging every month.
In many cases, an application's energy consumption can be significantly reduced,
while still meeting its SLA, if the application used a different less energy-intensive computing resource.

For example, FPGAs are often much more energy efficient than CPUs on the same computation.
However, for highly dynamic workloads with tight timing limits, CPUs are often used instead
because they can be quickly configured and/or reallocated as demand changes. We believe we can obtain
the best of both worlds by making it possible to run \functions in hybrid mode---using CPUs
to meet transient and short-term bursts with FPGAs used to meet the more stable and predictable 
portion of the workload. Because FPGAs, like CPUs, are at their peak energy efficiency at full utilization,
this means transparently scaling FPGAs up and down much like we do today for CPUs. 
To reduce engineering costs of maintaining multiple implementations, we aim to develop an 
intermediate representation (IR) that can be converted to run on a broad spectrum of accelerators (\eg similar to what TVM \cite{tvm} does for machine learning);
cloud customers will then be able to tradeoff between agility and energy efficiency as they see fit.

\begin{figure}[!t]
	\centering
	\includegraphics[scale=0.2]{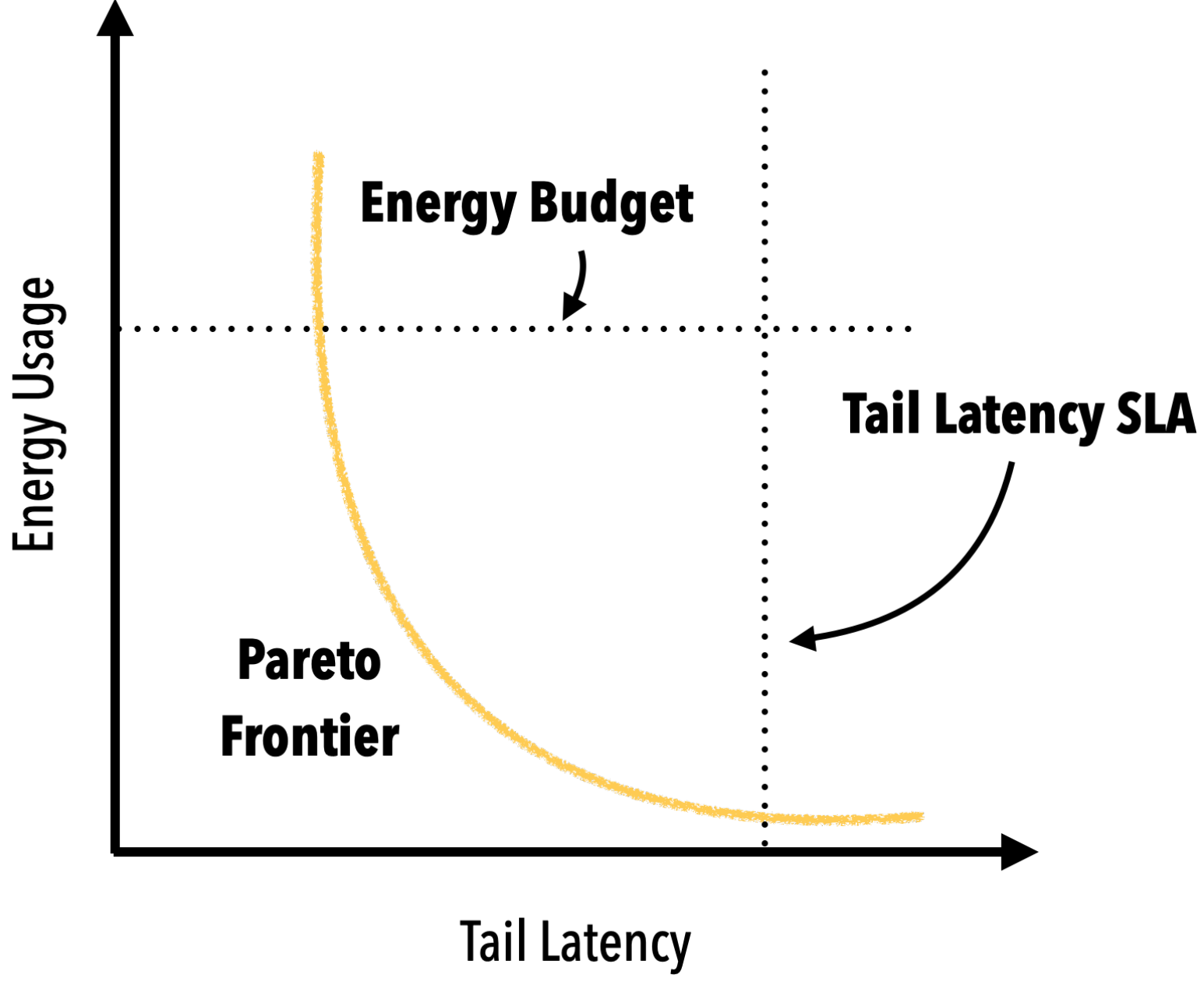}
	\caption{Pareto frontier of energy usage-vs-tail latency for interchangeable memory options.
    For example, DRAM and SSD are located at the opposite corners in this tradeoff space, but the relationships between alternatives may not always be linear.}
	\label{fig:pareto}
\end{figure}

\subsection{Interchangeable Memory}
Similar to interchangeable compute devices, DRAM, NVRAM, SSD, and HDD can all be interchanged to some degree:
while DRAM is volatile, in many use cases non-volatility is not a strict requirement.
Each offers a different operating point in the tradeoff between
energy efficiency and tail latency, as shown schematically in 
Figure~\ref{fig:pareto}. Even within a particular technology, 
there are often energy tradeoffs, such as in the choice between
single and multi-level cell encodings on SSDs.

Another trend is towards microsecond-scale networks, such as CXL and RDMA. 
This can allow memory resources to be more effectively disaggregated, reducing
both the cost and energy waste of resource stranding.
Combined with high-performance storage technologies, such 
as 3D XPoint (e.g., Intel Optane SSD~\cite{optane-ssd}) or SLC NAND (e.g., Samsung Z-SSD~\cite{z-ssd}) which offer microsecond-scale access times,
significant amounts of energy (and carbon) can be saved by shifting data
that is currently stored on DRAM to lower-power nearby storage.

We propose to design a \emph{general-purpose} system that
interchanges memory for lower power storage,
without affecting the application's SLAs while staying within an energy budget.
Such a system would need to automatically identify which data should sit in DRAM, 
and which part in storage, based on the \function's timeliness constraint, its read and write access patterns, and 
its access granularity. In addition, we can also employ intelligent 
caching and prefetching to mask reduced DRAM use \cite{leap}.

\subsection{Energy-Aware Scheduling}
\label{sec:scheduler}

So far, we have separately considered interchangeable compute and interchangeable memory resources.
For the most part, we have also assumed that the total energy consumption is given as a constraint for those optimizations. 
However, any realistic application requires both computation and storage.
We need to consider how to find the Pareto frontier of an application's energy-performance curve by co-optimizing both sets of interchangeable resources in a disaggregated environment, while taking energy sources and \function SLAs into account.

\begin{figure}[t]
	\centering
	\includegraphics[width=\columnwidth]{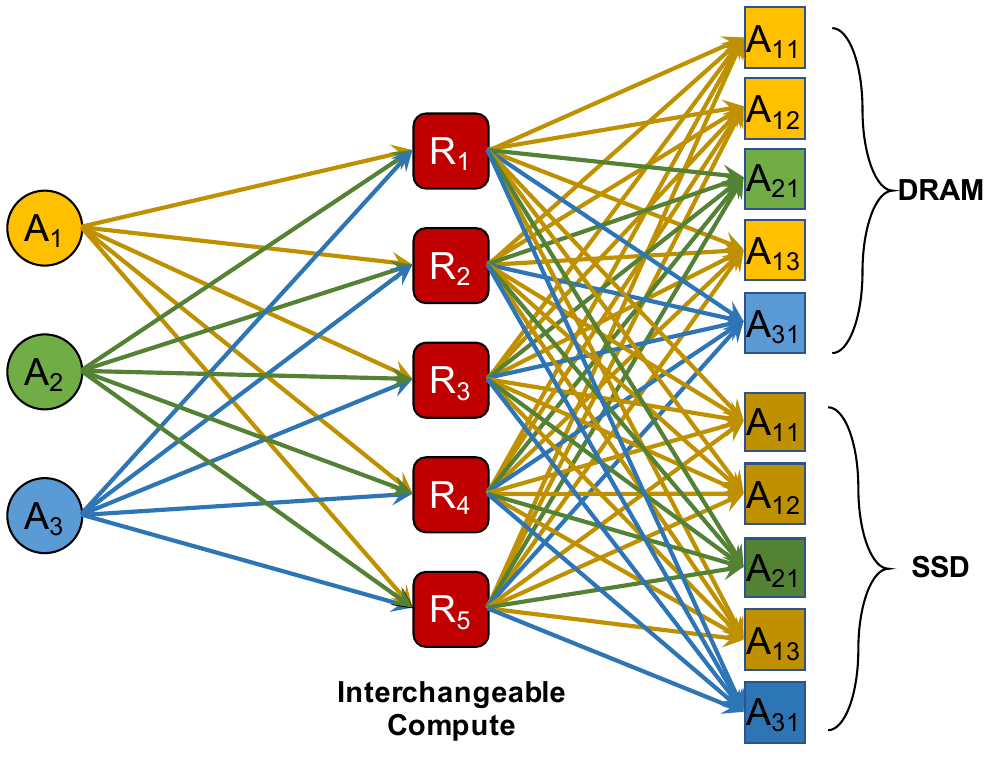}
	\caption{An example multi-commodity flow-based formulation for cost-performance optimization at time $t$.}
	\label{fig:energy-scheduling}
\end{figure}

Given that a \function can run on multiple interchangeable compute devices and the computation device may have choices to use one of the many storage mediums, one direction would be extending well-known multi-commodity flow-based resource allocation formulations \cite{mcf, isard2009quincy, allox:eurosys20} for determining the best combination of interchangeable resources to use.
Figure~\ref{fig:energy-scheduling} gives a simple example. 
There are \functions from three applications: $A_1$, $A_2$, and $A_3$ (three commodities with different colors), each of which can run on one of the five compute resources ($R_1 \ldots R_5$) with different -- already-profiled and known -- speedups.
At time $t$, each \function can read and write pertinent data (\eg, $A_1$ needs three objects $A_{11}$--$A_{13}$) from/to two interchangeable storage devices (the availability of data for reading can be captured by the presence/absence of edges between a compute device and corresponding data in that storage medium).
Now one can represent the problem of optimizing for total energy consumption for these simple \functions as the sum of all edge costs for each \function (with appropriate constraints to avoid oversubscribing each compute device) -- minimizing the total cost across all \functions will ensure that the overall energy consumption is minimized. 
By appropriately setting the costs of the edges and objective functions, we can consider trading off energy consumption for application performance and vice versa.
The primary challenge of such optimization-based approaches is the speed at which we can determine placements---a few microseconds may not be enough.
Approximation- and/or memoization-based are more likely to succeed.

What we highlighted so far deals only with assignments of \functions to interchangeable compute and memory/storage at a particular time instant. 
However, 
one-shot device assignment is just the beginning of the scheduling problem; we must 
also schedule \functions over time without violating their SLAs. 
The key here will likely be to take advantage of deadline-based and altruistic scheduling solutions \cite{grandl2016altruistic} to effectively leverage available slack.
We can consider dividing time into fixed-length scheduling windows, pack \functions with smaller slack within the current window, and push \functions with larger slack 
into future windows. This would maximize our ability to convert application
flexibility over timeliness into lower energy and carbon use. 

\section{Conclusion}
\label{sec:conclusion}
The end of Dennard scaling and the slowing of Moore's Law has led to an 
inflection point with respect to the impact the computing industry on 
the world's ecology.  Computing is still a small fraction of global 
energy use, but we can no longer count on automatic advances in the 
energy-efficiency of computing to compensate for the rapid upward spiral
in computing demand.  
To continue to reap the benefits of computing without endangering the planet, 
we need to treat energy efficiency as a first class design goal. 
The Treehouse project aims to address this challenge by building tools that help
application developers understand the implications of their design
decisions on energy and carbon use, by adapting interfaces to make timeliness
requirements explicit to allow for informed system-level tradeoffs of energy 
versus time, and by reducing the energy cost of commonly used abstractions.
More broadly, we believe that the systems software research community can 
and must play a constructive role in reducing the impact of computing 
on the planet, as we make the transition to abundant carbon-free energy 
over the next few decades.

\section*{Acknowledgments}
We would like to thank Simon Peter for suggesting Figure 1.
This work is supported by grants from the National Science Foundation (2104243, 2104292, 2104398, and 2104548), VMware, and Cisco Systems.

\balance
\bibliographystyle{abbrv}
\bibliography{paper}

\begin{thebibliography}{10}

\bibitem{aws-lambda}
{AWS Lambda}.
\newblock \url{https://aws.amazon.com/lambda/}.

\bibitem{optane-ssd}
{Intel Optane SSD 9 Series}.
\newblock
  \url{https://www.intel.com/content/www/us/en/products/memory-storage/solid-state-drives/consumer-ssds/optane-ssd-9-series.html}.

\bibitem{landuse}
National renewable energy laboratory: Land use by system technology.
\newblock \url{https://www.nrel.gov/analysis/tech-size.html}.

\bibitem{z-ssd}
{Samsung Z-SSD}.
\newblock \url{https://www.samsung.com/semiconductor/ssd/z-ssd/}.

\bibitem{ghg}
Greenhouse gas corporate accounting and reporting standard.
\newblock \url{https://ghgprotocol.org/corporate-standard}, 2021.

\bibitem{firecracker}
A.~Agache, M.~Brooker, A.~Iordache, A.~Liguori, R.~Neugebauer, P.~Piwonka, and
  D.~Popa.
\newblock Firecracker: Lightweight virtualization for serverless applications.
\newblock In {\em USENIX NSDI}, pages 419--434, 2020.

\bibitem{remote-regions}
M.~K. Aguilera, N.~Amit, I.~Calciu, X.~Deguillard, J.~Gandhi, S.~Novakovi{\'c},
  A.~Ramanathan, P.~Subrahmanyam, L.~Suresh, K.~Tati, R.~Venkatasubramanian,
  and M.~Wei.
\newblock {Remote Regions}: A simple abstraction for remote memory.
\newblock In {\em USENIX ATC}, 2018.

\bibitem{farmemory-throughput}
E.~Amaro, C.~Branner-Augmon, Z.~Luo, A.~Ousterhout, M.~K. Aguilera, A.~Panda,
  S.~Ratnasamy, and S.~Shenker.
\newblock Can far memory improve job throughput?
\newblock In {\em ACM EuroSys}, 2020.

\bibitem{ebs}
Amazon.
\newblock {Amazon Elastic Block Store}.
\newblock \url{https://aws.amazon.com/ebs/}.

\bibitem{s3}
Amazon.
\newblock {Amazon Web Services}.
\newblock \url{https://aws.amazon.com/s3/}.

\bibitem{energyprop}
L.~A. Barroso and U.~H{\"{o}}lzle.
\newblock The case for energy-proportional computing.
\newblock {\em Computer}, 40(12):33--37, 2007.

\bibitem{killermicroseconds}
L.~A. Barroso, M.~Marty, D.~A. Patterson, and P.~Ranganathan.
\newblock Attack of the killer microseconds.
\newblock {\em Commun. {ACM}}, 60(4):48--54, 2017.

\bibitem{IX}
A.~Belay, G.~Prekas, A.~Klimovic, S.~Grossman, C.~Kozyrakis, and E.~Bugnion.
\newblock {IX}: A protected dataplane operating system for high throughput and
  low latency.
\newblock In {\em USENIX OSDI}, pages 49--65, 2014.

\bibitem{tvm}
T.~Chen, T.~Moreau, Z.~Jiang, L.~Zheng, E.~Yan, H.~Shen, M.~Cowan, L.~Wang,
  Y.~Hu, L.~Ceze, et~al.
\newblock {TVM}: An automated end-to-end optimizing compiler for deep learning.
\newblock In {\em OSDI}, pages 578--594, 2018.

\bibitem{chen2012energy}
Y.~Chen, S.~Alspaugh, D.~Borthakur, and R.~Katz.
\newblock Energy efficiency for large-scale {MapReduce} workloads with
  significant interactive analysis.
\newblock In {\em ACM EuroSys}, pages 43--56, 2012.

\bibitem{chen2010compress}
Y.~Chen, A.~Ganapathi, and R.~H. Katz.
\newblock To compress or not to compress-compute vs. io tradeoffs for mapreduce
  energy efficiency.
\newblock In {\em ACM SIGCOMM Workshop on Green networking}, pages 23--28,
  2010.

\bibitem{cheng2015towards}
D.~Cheng, P.~Lama, C.~Jiang, and X.~Zhou.
\newblock Towards energy efficiency in heterogeneous {Hadoop} clusters by
  adaptive task assignment.
\newblock In {\em IEEE ICDCS}, pages 359--368, 2015.

\bibitem{enberg2019partition}
P.~Enberg, A.~Rao, and S.~Tarkoma.
\newblock Partition-aware packet steering using {XDP and eBPF} for improving
  application-level parallelism.
\newblock In {\em Proceedings of the 1st ACM CoNEXT Workshop on Emerging
  in-Network Computing Paradigms}, pages 27--33, 2019.

\bibitem{mcf}
S.~Even, A.~Itai, and A.~Shamir.
\newblock On the complexity of time table and multi-commodity flow problems.
\newblock In {\em 16th Annual Symposium on Foundations of Computer Science},
  pages 184--193. IEEE, 1975.

\bibitem{azure-smartnics}
D.~Firestone, A.~Putnam, S.~Mundkur, D.~Chiou, A.~Dabagh, M.~Andrewartha,
  H.~Angepat, V.~Bhanu, A.~Caulfield, E.~Chung, H.~K. Chandrappa,
  S.~Chaturmohta, M.~Humphrey, J.~Lavier, N.~Lam, F.~Liu, K.~Ovtcharov,
  J.~Padhye, G.~Popuri, S.~Raindel, T.~Sapre, M.~Shaw, G.~Silva, M.~Sivakumar,
  N.~Srivastava, A.~Verma, Q.~Zuhair, D.~Bansal, D.~Burger, K.~Vaid, D.~A.
  Maltz, and A.~Greenberg.
\newblock Azure accelerated networking: {SmartNICs} in the public cloud.
\newblock In {\em USENIX NSDI}, 2018.

\bibitem{xtrace}
R.~Fonseca, G.~Porter, R.~H. Katz, and S.~Shenker.
\newblock X-trace: A pervasive network tracing framework.
\newblock In {\em USENIX NSDI}, 2007.

\bibitem{app-performance-disagg-dc}
P.~X. Gao, A.~Narayan, S.~Karandikar, J.~Carreira, S.~Han, R.~Agarwal,
  S.~Ratnasamy, and S.~Shenker.
\newblock Network requirements for resource disaggregation.
\newblock In {\em OSDI}, 2016.

\bibitem{alibaba-disagg}
Y.~Gao, Q.~Li, L.~Tang, Y.~Xi, P.~Zhang, W.~Peng, B.~Li, Y.~Wu, S.~Liu, L.~Yan,
  F.~Feng, Y.~Zhuang, F.~Liu, P.~Liu, X.~Liu, Z.~Wu, J.~Wu, Z.~Cao, C.~Tian,
  J.~Wu, J.~Zhu, H.~Wang, D.~Cai, and J.~Wu.
\newblock When cloud storage meets {RDMA}.
\newblock In {\em {NSDI}}, 2021.

\bibitem{ghigoff2021bmc}
Y.~Ghigoff, J.~Sopena, K.~Lazri, A.~Blin, and G.~Muller.
\newblock Bmc: Accelerating memcached using safe in-kernel caching and
  pre-stack processing.
\newblock In {\em USENIX NSDI}, pages 487--501, 2021.

\bibitem{grandl2016altruistic}
R.~Grandl, M.~Chowdhury, A.~Akella, and G.~Ananthanarayanan.
\newblock Altruistic scheduling in multi-resource clusters.
\newblock In {\em USENIX OSDI}, pages 65--80, 2016.

\bibitem{infiniswap}
J.~Gu, Y.~Lee, Y.~Zhang, M.~Chowdhury, and K.~G. Shin.
\newblock Efficient memory disaggregation with {Infiniswap}.
\newblock In {\em NSDI}, 2017.

\bibitem{chasingcarbs}
U.~Gupta, Y.~G. Kim, S.~Lee, J.~Tse, H.-H.~S. Lee, G.-Y. Wei, D.~Brooks, and
  C.-J. Wu.
\newblock Chasing carbon: The elusive environmental footprint of computing,
  2020.

\bibitem{spdk}
{Intel Corporation}.
\newblock Storage performance development kit.
\newblock \url{http://www.spdk.io}.

\bibitem{perfiso}
C.~Iorgulescu, R.~Azimi, Y.~Kwon, S.~Elnikety, M.~Syamala, V.~R. Narasayya,
  H.~Herodotou, P.~Tomita, A.~Chen, J.~Zhang, and J.~Wang.
\newblock {PerfIso}: Performance isolation for commercial latency-sensitive
  services.
\newblock In H.~S. Gunawi and B.~Reed, editors, {\em USENIX ATC}, pages
  519--532, 2018.

\bibitem{isard2009quincy}
M.~Isard, V.~Prabhakaran, J.~Currey, U.~Wieder, K.~Talwar, and A.~Goldberg.
\newblock Quincy: fair scheduling for distributed computing clusters.
\newblock In {\em ACM SOSP}, pages 261--276, 2009.

\bibitem{jones2018stop}
N.~Jones.
\newblock How to stop data centres from gobbling up the world's electricity.
\newblock {\em Nature}, 561(7722):163--167, 2018.

\bibitem{profile_warehouse}
S.~Kanev, J.~P. Darago, K.~M. Hazelwood, P.~Ranganathan, T.~Moseley, G.~Wei,
  and D.~M. Brooks.
\newblock Profiling a warehouse-scale computer.
\newblock In D.~T. Marr and D.~H. Albonesi, editors, {\em ACM ISCA}, pages
  158--169, 2015.

\bibitem{flexnic}
A.~Kaufmann, S.~Peter, N.~K. Sharma, T.~Anderson, and A.~Krishnamurthy.
\newblock High performance packet processing with {FlexNIC}.
\newblock In {\em ACM ASPLOS}, pages 67--81, 2016.

\bibitem{kim2016data}
K.~Kim, F.~Yang, V.~M. Zavala, and A.~A. Chien.
\newblock Data centers as dispatchable loads to harness stranded power.
\newblock {\em IEEE Transactions on Sustainable Energy}, 8(1):208--218, 2016.

\bibitem{krioukov2011integrating}
A.~Krioukov, C.~Goebel, S.~Alspaugh, Y.~Chen, D.~E. Culler, and R.~H. Katz.
\newblock Integrating renewable energy using data analytics systems: Challenges
  and opportunities.
\newblock {\em IEEE Data Engineering Bulletin}, 34(1):3--11, 2011.

\bibitem{splinter}
C.~Kulkarni, S.~Moore, M.~Naqvi, T.~Zhang, R.~Ricci, and R.~Stutsman.
\newblock Splinter: Bare-metal extensions for multi-tenant low-latency storage.
\newblock In A.~C. Arpaci{-}Dusseau and G.~Voelker, editors, {\em USENIX OSDI},
  pages 627--643, 2018.

\bibitem{strata}
Y.~Kwon, H.~Fingler, T.~Hunt, S.~Peter, E.~Witchel, and T.~Anderson.
\newblock Strata: A cross media file system.
\newblock In {\em ACM SOSP}, Oct. 2017.

\bibitem{googledisagg}
A.~Lagar-Cavilla, J.~Ahn, S.~Souhlal, N.~Agarwal, R.~Burny, S.~Butt, J.~Chang,
  A.~Chaugule, N.~Deng, J.~Shahid, G.~Thelen, K.~A. Yurtsever, Y.~Zhao, and
  P.~Ranganathan.
\newblock Software-defined far memory in warehouse-scale computers.
\newblock In {\em ASPLOS}, 2019.

\bibitem{allox:eurosys20}
T.~N. Le, X.~Sun, M.~Chowdhury, and Z.~Liu.
\newblock {AlloX}: Compute allocation in hybrid clusters.
\newblock In {\em ACM EuroSys}, pages 31:1--31:16, 2020.

\bibitem{hydra:arxiv19}
Y.~Lee, H.~A. Maruf, M.~Chowdhury, A.~Cidon, and K.~G. Shin.
\newblock Mitigating the performance-efficiency tradeoff in resilient memory
  disaggregation.
\newblock {\em CoRR}, abs/1910.09727, Oct 2020.

\bibitem{secure}
J.~Li, S.~Miller, D.~Zhuo, A.~Chen, J.~Howell, and T.~Anderson.
\newblock An incremental path towards a safer os kernel.
\newblock In {\em Proceedings of the Workshop on Hot Topics in Operating
  Systems}, HotOS '21, page 183–190, 2021.

\bibitem{liu2011greening}
Z.~Liu, M.~Lin, A.~Wierman, S.~H. Low, and L.~L. Andrew.
\newblock Greening geographical load balancing.
\newblock {\em ACM SIGMETRICS Performance Evaluation Review}, 39(1):193--204,
  2011.

\bibitem{ma2020asymnvm}
T.~Ma, M.~Zhang, K.~Chen, Z.~Song, Y.~Wu, and X.~Qian.
\newblock {AsymNVM}: An efficient framework for implementing persistent data
  structures on asymmetric nvm architecture.
\newblock In {\em ACM ASPLOS}, pages 757--773, 2020.

\bibitem{snap}
M.~Marty, M.~de~Kruijf, J.~Adriaens, C.~Alfeld, S.~Bauer, C.~Contavalli,
  M.~Dalton, N.~Dukkipati, W.~C. Evans, S.~Gribble, N.~Kidd, R.~Kononov,
  G.~Kumar, C.~Mauer, E.~Musick, L.~E. Olson, E.~Rubow, M.~Ryan, K.~Springborn,
  P.~Turner, V.~Valancius, X.~Wang, and A.~Vahdat.
\newblock Snap: a microkernel approach to host networking.
\newblock In {\em ACM SOSP}, pages 399--413, 2019.

\bibitem{leap}
H.~A. Maruf and M.~Chowdhury.
\newblock {Effectively Prefetching Remote Memory with Leap}.
\newblock In {\em USENIX ATC}, 2020.

\bibitem{recalibrate-dc-energy}
E.~Masanet, A.~Shehabi, N.~Lei, S.~Smith, and J.~Koomey.
\newblock Recalibrating global data center energy-use estimates.
\newblock {\em Science}, 367(6481):984--986, 2020.

\bibitem{mashayekhy2014energy}
L.~Mashayekhy, M.~M. Nejad, D.~Grosu, Q.~Zhang, and W.~Shi.
\newblock Energy-aware scheduling of mapreduce jobs for big data applications.
\newblock {\em IEEE transactions on Parallel and distributed systems},
  26(10):2720--2733, 2014.

\bibitem{bento}
S.~Miller, K.~Zhang, M.~Chen, R.~Jennings, A.~Chen, D.~Zhuo, and T.~Anderson.
\newblock High velocity kernel file systems with {Bento}.
\newblock In {\em USENIX FAST}, pages 65--79, Feb. 2021.

\bibitem{shenango}
A.~Ousterhout, J.~Fried, J.~Behrens, A.~Belay, and H.~Balakrishnan.
\newblock Shenango: Achieving high {CPU} efficiency for latency-sensitive
  datacenter workloads.
\newblock In {\em USENIX NSDI}, pages 361--378, 2019.

\bibitem{avocados}
P.~Patel, K.~Lim, A.~Martinez, T.~Anderson, J.~Nelson, and I.~Zhang.
\newblock Fungible computing as a service.
\newblock https://treehouse-research.github.io/.

\bibitem{pearce2018energy}
F.~Pearce.
\newblock Energy hogs: can world’s huge data centers be made more efficient?
\newblock {\em Yale Environment}, 360, 2018.

\bibitem{pesce}
M.~Pesce.
\newblock {Cloud Computing's Coming Energy Crisis}.
\newblock {\em IEEE Spectrum}, 2021.

\bibitem{arrakis}
S.~Peter, J.~Li, I.~Zhang, D.~R.~K. Ports, D.~Woos, A.~Krishnamurthy,
  T.~Anderson, and T.~Roscoe.
\newblock Arrakis: The operating system is the control plane.
\newblock In {\em USENIX OSDI}, pages 1--16, 2014.

\bibitem{AIFM}
Z.~Ruan, M.~Schwarzkopf, M.~K. Aguilera, and A.~Belay.
\newblock {AIFM}: High-performance, application-integrated far memory.
\newblock In {\em USENIX OSDI}, pages 315--332, Nov. 2020.

\bibitem{shahrad2020serverless}
M.~Shahrad, R.~Fonseca, {\'I}.~Goiri, G.~Chaudhry, P.~Batum, J.~Cooke,
  E.~Laureano, C.~Tresness, M.~Russinovich, and R.~Bianchini.
\newblock Serverless in the wild: Characterizing and optimizing the serverless
  workload at a large cloud provider.
\newblock {\em arXiv preprint arXiv:2003.03423}, 2020.

\bibitem{LegoOS}
Y.~Shan, Y.~Huang, Y.~Chen, and Y.~Zhang.
\newblock Lego{OS}: A disseminated, distributed {OS} for hardware resource
  disaggregation.
\newblock In {\em USENIX OSDI}, 2018.

\bibitem{rdma-security}
A.~K. Simpson, A.~Szekeres, J.~Nelson, and I.~Zhang.
\newblock Securing {RDMA} for high-performance datacenter storage systems.
\newblock In {\em USENIX HotCloud}, July 2020.

\bibitem{dpdk}
{The Linux Foundation Projects}.
\newblock Data plane development kit.
\newblock \url{https://www.dpdk.org/}.

\bibitem{vahdat-sigcomm}
A.~Vahdat.
\newblock {Coming of Age in the Fifth Epoch of Distributed Computing}.
\newblock \url{https://www.youtube.com/watch?v=27zuReojDVw}.

\bibitem{behindserverless}
L.~Wang, M.~Li, Y.~Zhang, T.~Ristenpart, and M.~M. Swift.
\newblock Peeking behind the curtains of serverless platforms.
\newblock In {\em USENIX ATC}, pages 133--146, 2018.

\bibitem{kayak}
J.~You, J.~Wu, X.~Jin, and M.~Chowdhury.
\newblock Ship compute or ship data? why not both?
\newblock In {\em USENIX NSDI}, pages 633--651, 2021.

\bibitem{zhang2019i}
I.~Zhang, J.~Liu, A.~Austin, J.~Stephenson, and A.~Badam.
\newblock I’m not dead yet! the role of the operating system in a
  kernel-bypass era.
\newblock In {\em ACM HotOS}, April 2019.

\bibitem{justitia:nsdi22}
Y.~Zhang, Y.~Tan, B.~Stephens, and M.~Chowdhury.
\newblock {Justitia}: Software multi-tenancy in hardware kernel-bypass
  networks.
\newblock In {\em USENIX NSDI}, 2022.

\bibitem{bpf-storage}
Y.~Zhong, H.~Wang, Y.~J. Wu, A.~Cidon, R.~Stutsman, A.~Tai, and J.~Yang.
\newblock {BPF for Storage: An Exokernel-Inspired Approach}.
\newblock In {\em ACM HotOS}, 2021.

\end{thebibliography}

\end{document}